# Spin transfer nano-oscillators

Zhongming Zeng,[a] Giovanni Finocchio[b] and Hongwen Jiang[c]

[a]*Suzhou Institute of Nano-tech and Nano-bionics, Chinese Academy of Sciences, Ruoshui Road 398, Suzhou 215123, P. R. China*
[b]*Department of Electronic Engineering, Industrial Chemistry and Engineering, University of Messina, Messina 98166, Italy*
[c]*Department of Physics and Astronomy, University of California, Los Angeles, California 90095, United States.*

The use of spin transfer nano-oscillators (STNOs) to generate microwave signal in nanoscale devices have aroused tremendous and continuous research interest in recent years. Their key features are frequency tunability, nanoscale size, broad working temperature, and easy integration with standard silicon technology. In this feature article, we give an overview of recent developments and breakthroughs in the materials, geometry design and properties of STNOs. We focus in more depth on our latest advances in STNOs with perpendicular anisotropy showing a way to improve the output power of STNO towards the μW range. Challenges and perspectives of the STNOs that might be productive topics for future research were also briefly discussed.

## 1. Introduction

The magnetic states are classically controlled and manipulated through applied magnetic field. In 1996, John Slonczewski[1] and Luc Berger[2] predicted a different way to modify a magnetic configuration of nanomagnets based on the spin-transfer-torque (STT) effect from a spin-polarized current. Basically, the angular momentum carried by the spin-polarized current exerts a torque on the magnetization vector leading to either reversal or persistent precession. As result of the early experimental verification of STT effect,[3-6] there has aroused tremendous and continuous research interest.[7-15] The current-induced switching (reversal) offers promising applications in future magnetic random access memory (STT-MRAM). The current-induced magnetization precession enables magnetic nanostructures to be new type of high-frequency tuneable nanoscale oscillators, namely spin-transfer nano-oscillators (STNOs) or spin-torque nano-oscillators. The physics of STT has been previously reviewed[16] while a list of potential technological applications can be found in Ref. 17. This feature article is focused only on the description of all the issues related to the STNOs.

The STNOs have many intriguing advantages over a standard LC-tank voltage-controlled oscillator (VCO). Firstly, they are highly tuneable by bias current and magnetic field. While the frequency variation in VCO is only 20% compared to the carrier frequency, the oscillation frequency of the STNOs can be tuned by current over a range of several GHz and by magnetic field up to 40 GHz for particular configurations.[14] Secondly, STNOs are among the smallest microwave oscillators developed, over 50 times smaller than a standard VCO in complementary metal-oxide-semiconductor (CMOS) technology.[18] Thirdly, STNOs work over a broad range of temperatures and can be biased at low voltages (< 1.0 V). Finally, their simple structure is a key ingredient for on-chip realization. All those advantages make the STNOs promising for on chip-to-chip micro-wireless communications, small array transmitters, microwave sources for nanosensors, local on chip clocks for VLSI applications, and very high-density massively-parallel microwave signal processors.[19]

This feature article addresses the recent developments and breakthroughs in the STNO research. Some of the fundamental aspects of STNO have been thoroughly reviewed by S. E. Russek *et al.*[19] Other previously excellent review articles related to this field focuses on the point-contact devices[20] and on a theoretical analysis based on the universal model of non-linear oscillators.[21]

The STNOs can be roughly classified as the unpatterned point contact devices and patterned nanopillars according to the patterning geometry. There are different parameters describing the STNO performances, such as oscillator frequency, output power, linewidth, Q factor, power dissipation, phase noise, and robustness among them. Each parameter may be relevant in spite of the different device geometries. This review does not intend to exhaustively revise the "state of the art" on the optimization of all possible device performance characteristics. We rather focus on the most prominent remaining challenges and recent breakthroughs in current researches.

In Section 2, it is briefly introduced the fundamental working principles of STNOs. Section 3 describes the device configurations and a concise literature review. In Section 4, recent advances on output power and linewidth of STNO are discussed underlining our latest goals in presence of interface perpendicular anisotropy (IPA). Sections 5 and 6 remark on the modelling of STNOs and the non-autonomous dynamics, respectively. In the last section, the challenges and perspectives of the STNOs are also briefly discussed.

## 2. Operation principle

The simple configuration of a STNO consists of a relatively thick "fixed" magnetic layer, which serves as a polarizer, a nonmagnetic spacer, and a relatively thin magnetic "free" layer (see a sketch in Figure 1(a)). A dc current, spin-polarized from the polarizer, when large enough to transfer a sufficient STT to cancel out the intrinsic damping losses of the free layer, leads to a steady-state magnetization dynamics. The magnetoresistive (MR) effect converts the magnetization oscillation to a microwave voltage (Fig. 1(b)). The nonmagnetic spacer can be used, either a nonmagnetic metal (*e.g.* copper) or a thin dielectric (*e.g.* MgO), the corresponding structure is usually called a giant magnetoresistance (GMR) spin valve or a magnetic tunnel junction (MTJ). It is important to stress that in order that the STT becomes significant, high current densities are required (on the order of $10^7$ Acm$^{-2}$) achievable experimentally with applied currents of few milliamperes. For a spin valve device, it is

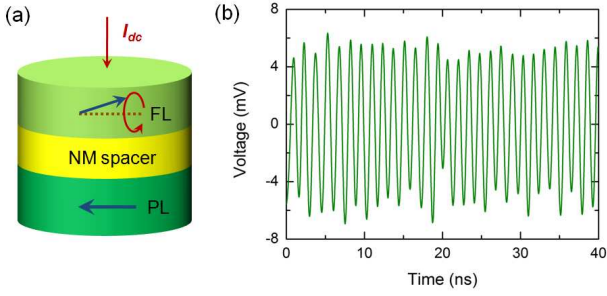

**Fig.1** (a) A STNO device consists of a "fixed" layer that serves as polarizer (PL) and a "free" layer (FL) whose magnetization is excited into steady state oscillations, NM spacer denotes non-magnetic layer, *i.e.* insulator or non-magnetic metal. (b) Voltage oscillations produced by steady-state precession of the magnetic free layer in a nanopillar sample, in response to the spin transfer torque from a 0.8 mA current. The sample had the core layer structure of IrMn/CoFe/Ru/CoFeB/MgO/CoFeB. The measurement was made at room temperature using a real-time oscilloscope.

trouble-free to apply large current to produce oscillations, but the output power is relatively low being the magnetoresistance signal small. On the other hand, for a MTJ an opposite scenario exists. A large output power can be delivered being the resistance change up to several hundred Ohms (see the paragraph about the output power in Section 4), but the maximum bias current is limited by the barrier breakdown voltage (~1.0 V). As consequence, the MTJ must have together to high tunnelling-MR (TMR) high transparency (low resistance-area product).[19]

## 3. Device configuration and brief review of literature

Seven years after the theoretical predictions[1, 2], Kiselev, Sankey and co-workers[12] performed a systematic experiment on microwave emissions in magnetic nanostructures. Over the last ten years, numerous theoretical and experimental studies have been made on STNOs. To date many different STNO configurations have been explored, which can be classified by the type of spacer layer (showed in Fig. 1(a)), the patterning geometry, and the equilibrium magnetic configurations. Two patterning geometries are used, one is to make "nanopillar" devices in which the free layer or both the free and the fixed layers are patterned to a desired cross section,[12] the other is to make electrical "point-contact" to an extended multilayer film (see Fig. 2(b)) referred as nano-contact.[14] Typical current densities for dynamical excitations are $10^7$-$10^8$ A/cm$^2$ in nano-contacts and $10^6$-$10^7$ A/cm$^2$ in nanopillars. Our recent work showed that the current density for excitations in nanopillars can be down to $10^6$ A/cm$^2$.[22] On the other hand, the nano-contact devices have the advantage of a narrow linewidth ($\Delta f$, typically 1~ 10 MHz) and a high quality factor Q (up to 18000, ref. 23) compared to the nanopillars. The origin of their different behaviour in term of dynamical response is related to the excitation of qualitative different modes, this aspect will be detailed discussed in the following paragraphs of this section.

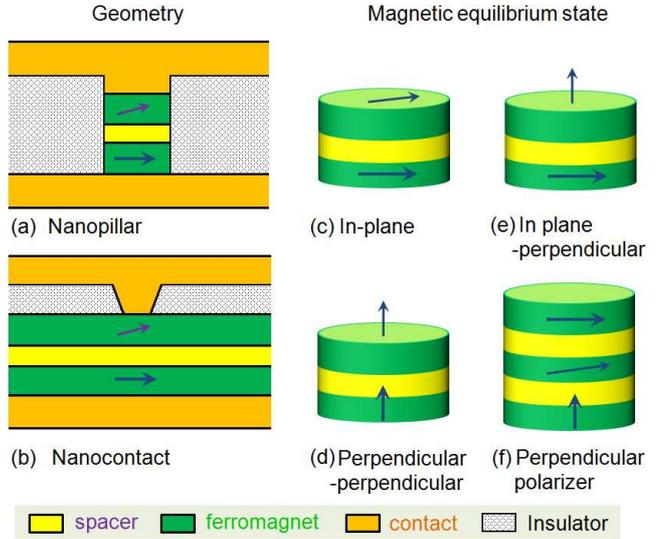

**Fig. 2** Different configurations of STNOs based on the type of the patterned geometry and the magnetic state.

The classification of a STNO depending on its magnetic configurations is very interesting. The most popular one is based on an in-plane configuration where the magnetization orientations of the free and fixed layer are designed to have an in-plane easy axis (see Fig. 2(c)).[12, 14] To produce microwave signal this structure generally requires a large static magnetic field applied a few degrees away from the magnetically easy axis of the free layer.[12] The second one is to make a "fully perpendicular" device in which both the free layer and polarizer are perpendicular to the sample plane as shown in Fig. 2(d).[24] The relative magnetic configuration in this structure is similar to that of the in-plane one, thus also requiring an external bias to produce microwave signal. The third one is to combine in-plane and out-of-plane magnetic configuration to build an in-plane-perpendicular device as shown in Fig. 2(e).[25, 26] This latter solution is very promising being an efficient configuration to excite large angular free layer precession leading to a large output power, even in the absence of a bias magnetic field. Based on the above configurations, hybrid devices can also be designed, for example, Fig. 2(f) presents a structure that utilizes a perpendicular polarizer and a planar free layer combined with a read-out layer.[27] However this read-out layer has a drawback of complicating the spin dependent transport in the STNO devices. In addition, in STNOs implemented with different ferromagnets for the free and the fixed layer (the "wavy" structure),[28] or with very different thickness (vortex structure),[29] has been measured microwave generation. While the former lacks in output signal strength, the latter is limited to maximum frequencies around 1 GHz and hence less attractive for practical application in microwave oscillators. In the following paragraphs, we would like to give a brief review of the current literature on experimental and theoretical results of the most common configurations of STNOs.

### 3.1 Nano-point contacts.

The first experimental observation of persistent magnetization dynamics has been reported by Tsoi and co-workers[3, 6] in point contact geometries with an out-of-plane bias field. Immediately after, Slonczewski identified those experimental excitations as exchange-dominated cylindrical spin wave modes (called 'Slonczewski mode') developing a spatially nonuniform linear theory of spin-wave excitations.[30] Those modes propagate radially out of the contact giving rise to the possibility to achieve the injection locking between the excitations of point-contacts localized at different position in the thin film nanomagnet.[31, 32] On the other hand, experiments by Rippard *et al.*[14] pointed out the excitation of magnetization precession also for in-plane bias fields. The linear theory developed in Ref. **Errore. Il segnalibro non è definito.** was not able to reproduce those latter experimental data. In fact, for in-plane magnetized magnetic film, it has been demonstrated analytically,[33] numerically by means of micromagnetic simulations,[34] and experimentally,[35] the excitation of a nonlinear localized spin wave mode (called "bullet mode"). The origin of these different kinds of excitation has been studied as function of the field angle systematically.[23, 34, 36] The key finding is that while the excitation threshold dependence on field angle of the bullet mode (from in-plane to out-of-plane) increases, the one of the Slonczewski mode decreases, this property gives rise to excitation of a bulled mode below a critical angle and the excitation of a Slonczewski mode above that critical angle. In addition, it has been demonstrated these two modes can be excited in a non-stationary way for a certain range of field angles and currents.[36] Micromagnetic simulations demonstrated that a large Oersted field is the key mechanism for the explanation of those latter experimental results.[35] Fig. 3 shows the oscillation frequency of the excited modes (Slonczewski (upper curve) and bullet (lower curve) modes) as function of the field angle. Fig. 3 left and right inset display the theoretically frequencies and an example of the experimental power spectral density where con be observed the two modes near 32.5 and 35 GHz.

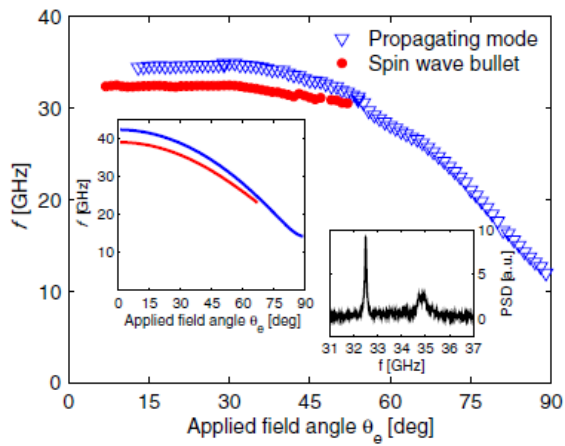

**Fig. 3** Measured frequencies of the observed spin wave modes as a function of the applied field angle. Left inset: theoretically calculated frequencies for the linear Slonczewski modes (upper curve) and for the bullet (lower curve) modes. Right inset: example power spectrum at 30 deg and a current of 14 mA. Reproduced with permission from ref. 36. Copyright 2010 American Physical Society.

The dynamical behaviour of nano-point contact at low bias fields (near zero field) changes completely. The excited modes are related to a vortex motion with circular trajectory around the

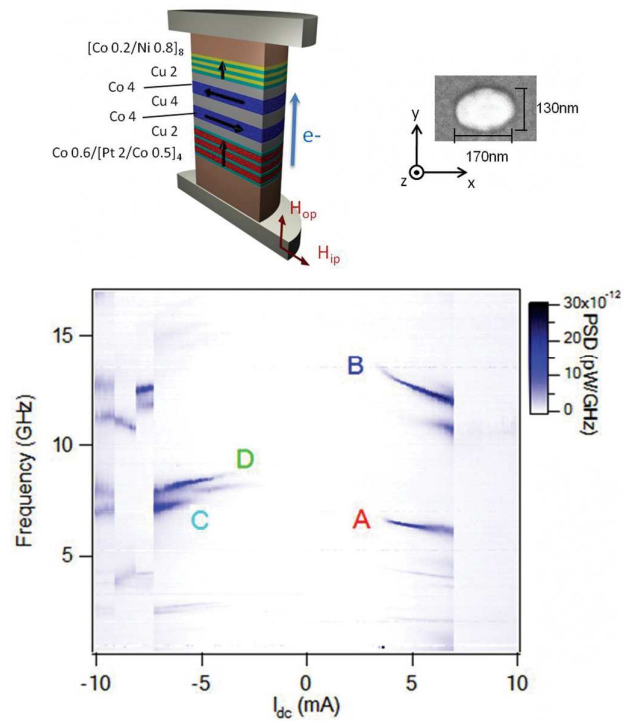

**Fig. 4** Top left: schematic of the nanopillar device geometry with two free layers and two perpendicular polarizers. Top right: top view of the nanopillar before depositing the top electrode, as viewed by a scanning electron microscope. Bottom: Measured microwave PSD as a function of the current bias at zero external field. Adapted with permission from ref. **Errore. Il segnalibro non è definito.**. Copyright 2012 American Physical Society.

injection site of the current[37] or with elliptical trajectory and continuous vortex polarity switching at very high current.[38] If the current injection is localized in the center of a spin-valve, rotating vortex-antivortex pair has been observed.[39] All the results are related to nano-contact geometries composed by ferromagnets (polarizer and free layers) with an in-plane easy axis. Recently, it has been also demonstrated excitation of persistent magnetization oscillation for nano-contacts where the easy axis of the free layer was out-of-plane.[25, 40]

A different idea to realize a nano-contact was proposed by Boone *et al.*[41] Instead to use a two dimensional extended film they use a ferromagnetic nanowire (extended only in one dimension) with the contact localized in the centre. This system seems promising for improving the locking distance between several different point-contacts being the losses due to the radiation of the wave below the contact smaller than the ones in the 2d point contacts.[42] The authors also think that the characterization, design and modeling of those devices will be a productive topic for future researches.

### 3.2 Spin-valves.

In 2003, it has been demonstrated experimentally the possibility to achieve persistent magnetization oscillation in confined systems, *i.e.* magnetic nanopillar "spin-valves" composed by Co/Cu/Co trilayer with elliptical cross section, by using a direct electrical measurement of the microwave frequency of the excited modes.[12] In a complete phase diagram of the magnetization dynamics as function of the bias field and current, several kinds of excited modes are reported. A successive paper, on the basis of micromagnetic simulations, identifies those excitations as non-uniform modes.[43] Only two years later, the final confirmation of the Slonczewski prediction is done with time domain measurements of the dynamics excited by spin-polarized current in a spin-valve with the polarizer exchange biased at 45° with respect to the easy axis of the free layer.[44] The main resulting advantage was the achievement of repeatable measurements. Complete analyses of those devices indicate the excitation of a uniform mode in a wide range of currents,[45, 46] and the possibility to achieve narrow linewidth for field applied near the in-plane hard axis.[47] This latter result has been explained by means of micromagnetic simulations as transition of the exited mode from non-uniform (field applied near the easy axis) to uniform (field applied near the hard in-plane axis).

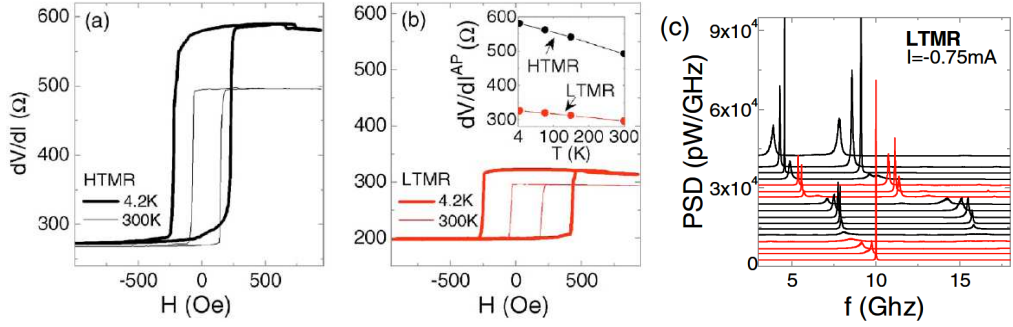

**Fig. 5** Field-dependent transfer curves for (a) HTMR and (b) LTMR elliptical devices at 4.2 and 300 K. The inset in (b) shows the evolution of the AP-state $dV/dI$ with temperature for both types of devices. Spectra (c) for an LTMR sample at constant current $I=-0.75$ mA and for fields of 0.2–1 kOe in steps of 50 Oe (from top to bottom), offset for clarity. Reproduced with permission from ref. **Errore. Il segnalibro non è definito.**. Copyright 2008 American Institute of Physics.

In those years, several studies on spin-valve based STNOs have been developed including perpendicular materials (the dynamics has been achieved experimentally only at very high fields and currents)[48] and half metallic materials with large spin polarization.[49, 50] Another solution was the use of an in-plane free-layer with an out-of-plane polarizer.[27, 51] In spin-valves with one of the two layers thick enough (60 nm) was observed the first excitation of vortex dynamics driven by spin-transfer torque.[29] In that system, micromagnetic simulations demonstrated that both magnetic layers were involved in the magnetization dynamics.[52, 53] A very promising solution for practical applications is the use of spin-valves with two in-plane free layers and two out-of-plane polarizers;[54] see Fig. 4 (top left) for a sketch of that nanopillar and Fig. 4 (top right) for a scanning electron microscope image of top view of the nanopillar before depositing the top electrode. The main result achieved in this configuration is the largest oscillation frequency at zero bias field (>5GHz), for instance, Fig. 4 (bottom part) displays the measured microwave power spectral densities PSDs as a function of the bias current without magnetic field. Micromagnetic simulations indicate that the dynamics of the two free layers is phase locked and the coupling mechanisms (*i.e.* magnetic dipolar field and the back torque) are both important in reproducing the experimental behavior.

### 3.3 Magnetic tunnel junctions.

The first roughly report of measurements of persistent magnetization dynamic in MTJs with AlO tunnel barrier was presented by Fuchs *et al* in 2003.[55] However, only after the discovery of a giant TMR at room temperature achieved with MgO tunnel barrier,[56, 57] it was possible to perform a systematic characterization of the dynamical properties of MTJ-based STNOs. The first interesting result was obtained in 2006, a pronounced narrow peak (linewidth around 21 MHz) at the frequency of around 7 GHz was measured.[58] Other measurements[59] reported that two groups of MTJ STNOs (see Fig. 5), *i.e.* high resistance with high TMR ratio (labeled HTMR) and low resistance with low TMR ratio (labeled LTMR), exhibited distinct dynamic behaviors. In particular, the LTMR devices exhibited narrow linewidths (< 10 MHz) as shown in Fig. 5(c), while HTMR devices can deliver large output power, as also supported from recent studies.[60, 61]

Vortex dynamics has been also measured in MTJs (see Fig. 6), the key difference with respect to the vortex dynamics in spin-valve is the presence of an exchange biased polarizer resulting in a dynamical behavior present only in one ferromagnet.[62]

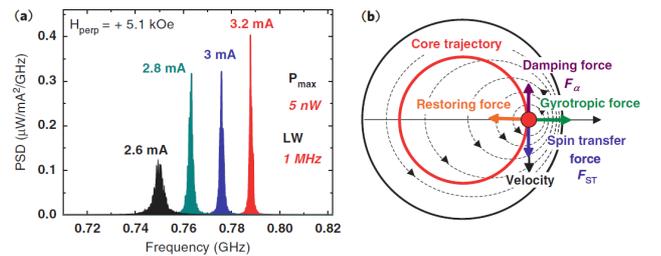

**Fig. 6** Large microwave emission spectra associated with vortex oscillations. (a) Power spectral densities (PSD) normalized by $I^2$, obtained for $I = 2.6, 2.8, 3$ and $3.2$ mA with $H_{perp} = 5.1$ kOe. (b) Schematic of the different forces having an effect on the vortex core. Reproduced with permission from ref. 62. Copyright 2010 Nature Publishing Group.

Another milestone in the design of STNOs was the discovery of perpendicular MTJs by Ikeda *et al.*[63] It gave rise to a number of experiments where, by controlling the trade-off between shape anisotropy and IPA, it is possible to define the easy axis of either polarizer and free layer. In particular, for STNO applications, the IPA can be used to reduce the out-of-plane demagnetizing field while maintaining the orientation of both the two magnetizations in the film plane,[Errore. Il segnalibro non è definito., 64] or by maintaining the polarizer in-plane and the free layer tilted[65] or out-of-plane.[26] Those latter results are very promising to design high performance STNOs where all the parameters discussed later in the text can be optimized at the same time.

## 4. Output power and linewidth

The output power and linewidth are the two most important parameters for a STNO. Although large numbers of experiments have been made to push the technology beyond the mere proof of concept, the important bottlenecks in STNO technology lies in the enhancement of its output power and simultaneously the improvement of the linewidth. The difficulty of attaining high power and narrow linewidth STNO arises not only in device fabrication, but also in the choice of the material and the design. In line of principle, one can improve the performance of a single STNO configuration,[22] and then further advance to the practical level by means of synchronization of an array of that STNO (see the excellent previous reviews 19, 21).[31, 32, 66, 67] However, this advance has not been made yet.

### 4.1 Output power

We would like to first look at the key factors to determine the output power of a single STNO. It is well known that the dc current ($I$) flowing perpendicular to the layers of a STNO excites magnetization oscillations in the free layer, which give rise to a temporal variation in the resistance due to MR effect, $R(t) = R + \Delta R/2 \times cos(\omega t)$, where $R$ is approximately equal to the dc resistance and $\Delta R$ is the oscillation amplitude of the resistance induced by $I$. The temporal variation of the resistance generates a microwave voltage $V(t) = I \times R(t)$. Thus the output power delivered from the STNO to a load with impedance $R_L$ is approximately given by[19]:

$$P_{out} = \frac{V_{out}^2}{2R_L} = \frac{I^2}{8} \frac{\Delta R^2 R_L}{(R+R_L)^2}, \quad (1)$$

From this equation it is evident that maximizing the output power requires one to maximize the $\Delta R$ value and to optimize the impedance matching ratio $R/R_L$. It is worthwhile to note that $\Delta R = A_m \times MR$ ($A_m$ is related to the amplitude of magnetization oscillation, $MR = (R_{AP}-R_P)/R_P$, $R_{AP}$ and $R_P$ are the resistances when the magnetizations between the fixed and free layer are aligned anti-parallel and parallel respectively). Thus the output power is approximately proportional to the *MR* effect, the oscillation amplitude of magnetization and the impedance match ratio $R/R_L$. For example by considering typical experimental values, for a spin-valve with MR = (0.1–10%), $I = 1$ mA, $R = 5$ Ω, $\Delta R = 0.5$ Ω, and $R_L = 50$ Ω, the resulting $P_{out} \approx 0.5$ nW, while for an MTJ with a large MR (> 50%), assuming $I = 1$ mA, $R = 450$ Ω, $\Delta R = 150$ Ω, and $R_L = 50$ Ω into the above equation, the $P_{out}$ will be up to 0.5 μW. Recent experiments demonstrated that MgO-based STNOs with large TMR ratio are able to deliver this large power, due to the excitation of a uniform mode where the magnetization oscillates around the equilibrium configuration of the magnetization as also indicated by micromagnetic simulations.[22, 61] Even so, there is still room to boost the output power for MTJ STNO through the achievement of the large-amplitude oscillations[46, 68, 69] being the oscillation amplitude value in the above reports smaller than the corresponding full MR change.

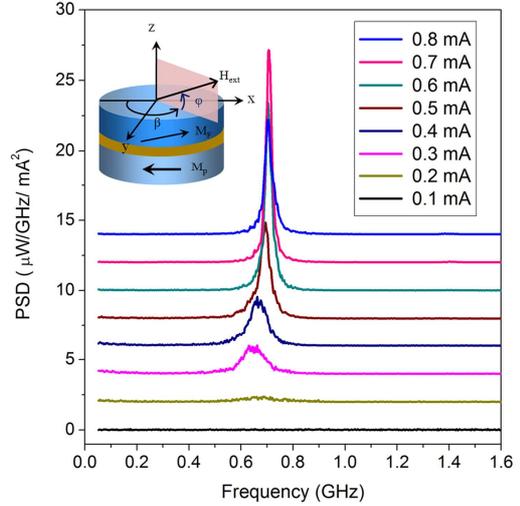

**Fig. 7** Microwave emission spectra measured at positive currents between 0.1 mA and 0.8 mA with 0.1 mA steps. The power spectral densities (PSD) normalized by $I^2$ have been offset by 2 μW/GHz/mA² for clarity. The external magnetic field is applied at out-of-plane direction (*i.e.* $\beta$ =120° & $\varphi$ = 40.4°) with an amplitude of 264 Oe. Inset: Definition of the coordinate system. Here $\beta$ is the angle between the external magnetic field $H_{ext}$ and the reference layer in the x-y plane, and $\varphi$ is the angle with respect to the x-y plane. Adapted with permission from ref. 22. Copyright 2012 American Chemical Society.

Maximizing $\Delta R$ value requires both large MR and large oscillation amplitude. Slavin and Tiberkevich[21] theoretically found that the oscillation amplitude and the resulting power increase with $I/I_c$, where $I_c$ is the critical current for the onset of STT-induced microwave emission, even if this expression has been derived for spin-valves and for currents near the $I_c$, our experience on MTJ experimental data permit to say that this expression is valid for almost the whole dynamical range.[21] However, in MTJ STNOs the maximum bias current is limited by the barrier breakdown voltage (~ 1 V). In many cases this does not allow for the application of sufficiently large bias currents needed to induce large-amplitude oscillations. Moreover, at large bias currents, the MR ratio decreases due to spin excitations localized at the interfaces between the magnetic electrodes and the tunnel barrier.[70] The solution is thus to reduce $I_c$, the $I_c$ or $J_c$ (here $J_c$ the critical current density is $I_c/A$, $A$ is the device area) values in the previously reports are larger than $10^6$ A/cm². Very recently, our works[22, 26] demonstrated that the $I_c$ value can be significantly reduced by introducing the IPA the free layer, which can partially cancel the effect of the demagnetizing field.[71-74] The STNO utilizing perpendicular anisotropy exhibits simultaneously highest output power 0.28 µW (> 0.95 µW if delivered to the matched load) and a linewidth smaller than 25 MHz (comparable to that of LTMR STNOs[75, 76]) in the same device, as well as the suppression of the secondary oscillation modes. Fig. 7 shows typical microwave emission spectra for STNO devices with IPA. The data show a very large-amplitude oscillation (ΔR corresponds to approximately 85% of the total static resistance change due to MR effect) as confirmed by micromagnetic simulations. We stress the fact that previously reported results on STNOs with large power exhibited large linewidth and multiple peaks.[60] This is also different from the reported STNO with Fe-rich free layer,[64] where the observation of the large power (up to 0.175 µW) requires a large bias current ($I = 1.9$ mA) and a large external magnetic field ($H = 4500$ Oe) which is not convenient for applications. The performance of our STNO will expect to improve by phase-locking or used as basis in a synchronized STNO array.[66, 67] For instance, by considering 10 MTJ STNOs in series synchronized, we could get out at least 28 µW. We underline once again that the STNO with a built-in perpendicular anisotropy is promising toward developing high-power STNOs for practical applications.

### 4.2 Linewidth

For any oscillators, the linewidth (full width at half maximum of the power spectra, $\Delta f$) is a measure of the phase noise, which is one of the most important figures of merit.[77] Fig. 8 shows a power spectra of a STNO with $f = 0.71$ GHz, and $\Delta f = 25$ MHz. Differently from linear or quasi-linear oscillators where the phase noise is decoupled from the power noise, in STNOs exists an intrinsic coupling between oscillation amplitude and phase due to the dependence of the effective field on the magnetization. This coupling gives rise to an additional contribution to the phase noise coming from a renormalization of the power noise via the non-linear frequency shift ($N$).[78]

In general, the linewidth of STNOs strongly depends on both device geometry and materials and the operation conditions. The linewidth for a single nano-contact device can vary from a few

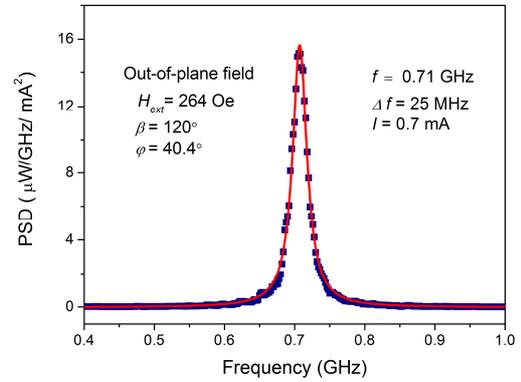

**Fig.8** An example of MTJ-based nanopillar STNO spectra showing a narrow linewidth at room temperature. The STNO geometry is the same as that in ref. 22, the elliptical pillar size is approximately 170 nm × 60 nm.

MHz to 100 MHz as the current and applied magnetic fields are varied,[14, 20, 25, 79] while linewidths for nano-pillars can vary from several tens MHz[22, 58, 59, 61, 75, 76] to the GHz level.[12, 23, 60, 61, 64] One of remaining challenges in the STNO design is the achievement of a narrow linewidth and a large output power at the same time in order to utilize them for practical applications.

First of all, the linewidth decreases as the emitted power $p_0$ increases.[80] As consequence small-amplitude dynamics are strongly influenced by thermal fluctuations,[81] while as confirmed by our previous experiments large oscillation amplitude induces a narrower linewidth.[22]

One important result from fundamental point of view has been achieved by Thadani *et al*[47] in spin-valves, they measured a strong variation of the oscillator linewidth as function of the in-plane field angle. This result was also confirmed for MTJ-based STNOs by Mizushima *et al*.[82] The theoretical explanation based on the universal model of non-linear oscillators indicates a minimum of the value of the coupling between phase and power $N$ corresponding to the minimum of the linewidth.[83] On the other hand, micromagnetic simulations also identify this variation as due to the transition from the excitation of a non-uniform to

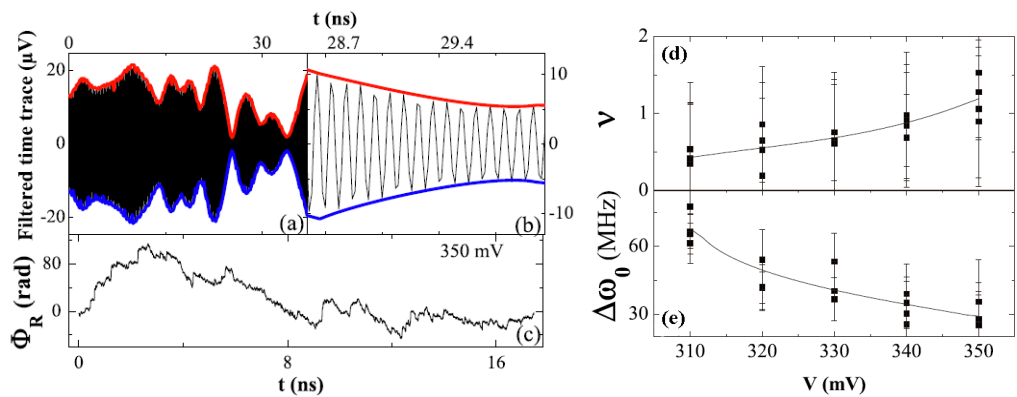

**Fig. 9** (a) Time-resolved voltage trace at $V=350$ mV and $H=-32$ mT in a 37 ns window with the time-varying amplitude A(t) of $x_a$(t) and -A(t) as the upper and lower envelope traces. (b) A 2.6 ns window zoom of the previous trace. (c) Deduced time-drift of the phase deviation $\Phi_R$(t). (d) The nonlinear coefficient ν and (e) the linear linewidth $\Delta\omega_0$ as a function of applied bias voltage. Lines indicate the mean values for each voltage. Adapted with permission from ref. 86. Copyright 2010 American Institute of Physics.

uniform mode.[84]

Those studies showed the first way to reduce the linewidth. It is based on the identification of a working point where $N$ is as low as possible. Several procedures have been presented to experimentally measure $N$. The first study has been presented by Keller *et al*[85], while Bianchini *et al*[86] developed a procedure for a direct experimental measure of this coupling by means of time domain data, see the plot of $\nu$ and linewidth as function of voltage in Fig. 9 d and e, respectively, as computed from the time domain traces (see for example Fig. 9 left). In general, the analysis of the time-domain measurements allows to separate the linear and nonlinear contributions to the phase noise, being the nonlinear contribution due to the power-phase coupling.[87]

For high current regime, the linewidth is not only related to the thermal fluctuations, but an additional intrinsic linewidth is present due to the chaotization of the magnetization trajectory[20, 46] as also observed in recent experimental works.[75, 88-90]

In addition, for real STNO devices the micromagnetic effects caused by devices edges, defects or inhomogeneous current injections are sources of nonuniform precession and thereby of linewidth broadening.

A second way to reduce the linewidth is the excitation of a vortex mode, where the linewidth can also exhibit sub-MHz linewidth with a sub-GHz oscillator frequency.[29, 62, 91] While micromagnetic simulations can reproduce also quantitatively the experimental linewidth,[53] it is worth to note that the theory developed in Ref. 78 cannot be directly applied and research towards generalizing that model will be very useful for a better understanding of the dynamical properties of the vortex based STNOs.

However, the synchronization of an array of STNOs is the most promising way to a sensibly reduction of the linewidth. Experiments indicate linewidths at least one order of magnitude smaller for synchronized STNOs compared to the same STNO unsincronized.[31,32,92] Experimental progress on this matter remains modest after the pioneering works[31,32] and the demonstration of an array of 4 STNO synchronization, and will be a very hot research topic in the next future.[92]

## 5. Modeling of spin transfer nano-oscillators

This section is devoted to present an overview of the modelling approaches used for the description of STNOs. In particular, there will be discussed the key differences to take into account in order to reproduce the magnetization dynamics in spin-valves, MTJs and point contact geometries in the micromagnetic framework. We underline that the numerical results already cited in the text can be achieved within the models described in the following paragraphs.

### 5.1 Full micromagnetic simulations.

The physical model that describes the dynamical behaviour of STNO is the micromagnetism. It has been used with success to reproduce, understand, and predict the experimental features of STNOs. The basic relationship is the Landau-Lifshitz-Gilbert equation with an additional term due to the Slonczewski spin-transfer torque (LLGS)[1]:

$$\frac{d\mathbf{m}}{d\tau} = -(\mathbf{m} \times \mathbf{h}_{\text{eff}}) + \alpha \mathbf{m} \times \frac{d\mathbf{m}}{d\tau} - \frac{g|\mu_B|j}{e\gamma_0 M_s^2 d}\varepsilon(\mathbf{m},\mathbf{m}_p)\mathbf{m} \times (\mathbf{m} \times \mathbf{m}_p) \quad (2)$$

where $g$ is the gyromagnetic splitting factor, $\gamma_0$ is the gyromagnetic ratio, $\mu_B$ is the Bohr magneton, $\alpha$ is the Gilbert damping, $j$ is the current density, $d$ is the thickness of the free layer, $e$ is the electron charge, $\mathbf{m}$ is the normalized magnetization of the free layer, $\mathbf{m_P}$ is the normalized magnetization of the polarizer, $M_S$ is the saturation magnetization, $d\tau$ is the dimensionless time step in unit of $\gamma_0 M_S$, and $\varepsilon(\mathbf{m},\mathbf{m_P})$ characterizes the angular dependence of the spin torque polarizing function. By convention, positive current polarity corresponds to electron flow from the free to the pinned layer of the spin valve. $\mathbf{h}_{\text{eff}}$ is the dimensionless effective field given by the standard micromagnetic contributions from exchange, anisotropy, external, demagnetizing fields, and in addition the Oersted field due to the current and the possible dipolar coupling between the polarizer and the free layer. The first term of the equation represents the conservative dynamics of the magnetization vector around the effective field, the second term is related to the intrinsic magnetic losses while the last term can be seen as positive damping term which furnishes energy to the system by means of the current. In general, the integration of eqn (2) is not trivial and it should be performed numerically. A complete numerical treatment of the problem and a review of the state of the art of numerical techniques to solve the LLGS equation are given in ref. 93. Here, we just point out some brief remarks in using eqn (2) for STNO modeling. For spin-valve based STNOs the current density is considered uniform in the whole cross section of the device, the polarizing function has been derived by Slonczewski firstly in the ballistic regime[1]:

$$\varepsilon(\mathbf{m},\mathbf{m}_p) = \left[-4 + (1+\eta)^3(3+\cos(\theta))/4\eta^{3/2}\right]^{-1} \quad (3)$$

where $\eta$ is the spin torque efficiency and $\theta$ is the angle between $\mathbf{m}$ and $\mathbf{m_P}$, and secondly using the dual channel model for symmetric spin-valves correlating directly the polarization function to the magnetoresistance $r$ <sup>Errore. Il segnalibro non è definito.</sup>:

$$\begin{cases} \varepsilon(\mathbf{m},\mathbf{m}_p) = 0.5P\Lambda^2/(1+\Lambda^2+(1-\Lambda^2)\cos(\theta)) \\ r(\mathbf{m},\mathbf{m}_p) = (1-\cos^2(\theta/2))/(1+\chi\cos^2(\theta/2)) \end{cases} \quad (4)$$

where $\chi$ is the GMR asymmetry parameter, $P$ is the current spin-polarization factor, and $\Lambda^2 = \chi + 1$. For complex configurations such as vortex or two free layer STNOs, it is necessary to simulate the coupled behavior of two ferromagnets in order to understand the experimental behavior.[53, 54] In that case, the LLGS equation to integrate for each ferromagnet is:

$$\begin{cases} \dfrac{d\mathbf{m}}{d\tau} = -(\mathbf{m} \times \mathbf{h}_{\text{eff}}) + \alpha \mathbf{m} \times \dfrac{d\mathbf{m}}{d\tau} - T(\mathbf{m},\mathbf{m}_p) \\ \dfrac{d\mathbf{m}_p}{d\tau} = -(\mathbf{m}_p \times \mathbf{h}_{\text{p-eff}}) + \alpha_p \mathbf{m}_p \times \dfrac{d\mathbf{m}_p}{d\tau} - T(\mathbf{m}_p,\mathbf{m}) \end{cases} \quad (5)$$

Eqn (5) takes into account both coupling mechanisms, the magnetostatic field computed by solving the magnetostatic problem for the whole spin-valve (ferromagnet/normal metal/ferromagnet) and the spin-transfer torque:

$$\mathbf{T}(\mathbf{m},\mathbf{m}_p) = \frac{g\mu_B j}{e\gamma_0 dM_s^2}\begin{cases}\mathbf{m}\times[\varepsilon(\mathbf{m},\mathbf{m}_p)(\mathbf{m}\times\mathbf{m}_p)] \\ -\mathbf{m}_p\times[\varepsilon(\mathbf{m}_p,\mathbf{m})(\mathbf{m}_p\times\mathbf{m})]\end{cases} \quad (6)$$

where $\mathbf{m_P}$ is not considered pinned and the polarizing function are the same

$$\varepsilon(\mathbf{m},\mathbf{m}_p) = \varepsilon(\mathbf{m}_p,\mathbf{m}) \quad (7)$$

To model the nano-point contact geometries there are two additional aspects to consider. Firstly, the current is spatially dependent in the sense that is locally injected in the ferromagnet via a nano-point contact, to reproduce the experimental data it has been demonstrated that can be considered zero outside the current injection site with a good approximation.[94] Secondly, the dimension of the free layer in nano-point contacts is very large (more than 2 µm × 2 µm) and because the discretization mesh is limited by the exchange length (of the order of 5-6 nm) the computational cost in term of both memory and time is very high. One approach to solve this problem was the reduction of the computational area (800×800 nm$^2$) by including some *ad hoc* boundary conditions[94-96] at the end of the computational region to avoid spurious effects due to possible reflection of the spin-waves. However, with the development of GPU-based parallel micromagnetic solvers, the computational region can be increased almost up to the real dimensions of the experimental free layer.[93]

The modeling of MTJs is the more complicated and at the same time it will be very important in the next future due to the recent perspective to realize commercial spintronic devices based on MTJs. The first difference with spin-valves and point contact geometries is the presence of an additional important field-like (or out-of-plane) torque term which depends on the bias voltage.[97-100] The total STT is given by:

$$T_{IP} + T_{OP} = \frac{g|\mu_B|J(\mathbf{m},\mathbf{m}_p)}{|e|\gamma_0 M_s^2 d}\varepsilon_T(\mathbf{m},\mathbf{m}_p)[\mathbf{m}\times(\mathbf{m}\times\mathbf{m}_p)] - q(V)(\mathbf{m}\times\mathbf{m}_p) \quad (8)$$

where the $T_{IP}$ and $T_{OP}$ are the in-plane and the out-of-plane torque respectively. $q(V)$ is the coefficient that takes into account the voltage dependence of the $T_{OP}$.[101] $\varepsilon_T$ represents the angular dependence of the spin torque polarizing function for MTJ as derived by Slonczewski:[102,103]

$$\varepsilon_T(\mathbf{m},\mathbf{m}_p) = 2\eta_T(1+\eta_T^2\cos(\theta))^{-1} \quad (9)$$

where $\eta_T$ is the spin torque efficiency for the MTJ stack. The last remark is related to the dependence of the current density distribution on the magnetization state. In fact, for high TMR MTJ and in presence of magnetic domains the local resistance can change significantly from domain to domain. A simple approximation to take into account this effect is the use of parallel multi-channel resistance.[104]

### 5.2 Simplified approaches, an overview.

Simpler theoretical approaches based on macrospin approximation[105] or universal model of non-linear oscillators (UMNO)[21] have been used to describe uniform dynamics. Sun[71] in 2000 solved, in the macrospin approximation, the LLGS equation achieving for the first time persistent magnetization oscillations numerically. Analytical solutions of the LLGS equation based on Melnikov theory can be found when both intrinsic damping and spin transfer torque can be treated as perturbations of the conservative dynamics.[105] The key results achieved in that framework are the existence of limit cycle (self-oscillation) of dynamics where the critical current represents a Hop bifurcation, and the fact that the chaos is precluded. Extension of this theory has been also used to predict the non-autonomous behavior of STNOs.[106,107] **Errore. Il segnalibro non**

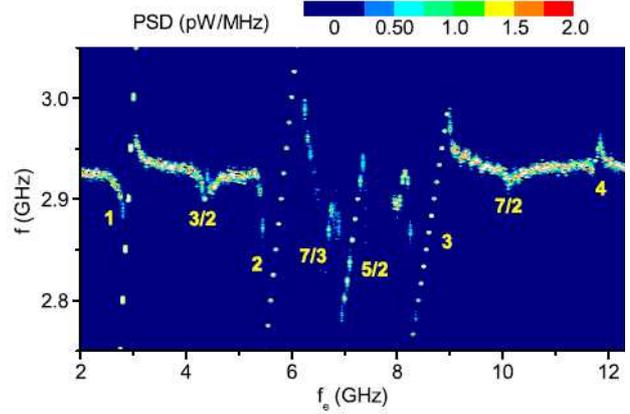

**Fig. 10** Power spectral density (PSD) spectra as function of microwave frequency. The frequency of the excited mode with larger power is well identified. The scale indicates the power spectral density (PSD) of the spectral signal. Reproduced with permission from ref. 109**Errore. Il segnalibro non è definito.**. Copyright 2010 American Physical Society.

**è definito.**

A more general approach based on the use of the UMNO has been proposed by Slavin and Tiberkevich[21] which can be used when a single mode is excited, the system is weakly non-conservative, and the excitation of the mode is critical. They demonstrated that the equation of the UMNO can be directly derived from the LLGS equation in a particular configuration.

$$\frac{dc(t)}{dt}+i\omega(|c|^2)c(t)+\Gamma_+(|c|^2)c(t)-\Gamma_-(|c|^2)c(t) = f(t) \quad (10)$$

In the eqn (10), $c(t)$ is the complex amplitude of the self-oscillation which measures both power $p=|c|^2$ and phase $\phi = \arg(c)$ of the oscillator. $\omega(|c|^2)$ is the resonant frequency, $\Gamma_+(|c|^2)$ and $\Gamma_-(|c|^2)$ are the positive and negative damping respectively. $f(t)$ represents the interaction of the oscillator with the rest of the world. One difficulty in the use of the UMNO is the identification of the model parameters. This can be directly done from the experimental data[108,109] or micromagnetic simulations[110] in the autonomous regime and then the model can be used to predict the behaviour in the non-autonomous regime or in presence of the thermal effects (prediction of the linewidth).

In presence of the excitation of non-uniform modes a different formalism should be used, see for example ref. 111 for the case of vortex-antivortex pair dynamics and a generalized Thiele equation[62,112] for the vortex dynamics.

## 6. Non-autonomous dynamics

The rich non-autonomous behaviour of the STNO (microwave signal together to a bias current) makes the system very attractive also from the fundamental point of view. Those properties are related to the intrinsic strong non-linearities of the system being the output power and the oscillator phase of the STNO coupled. This dependence is due to the dependence of the effective magnetic field on the magnetization itself. The phase locking at first harmonics of the oscillator frequency due to a microwave current has been experimentally studied in point-contact geometries[113] and in nanopillars.[114-117] In both cases the locking region was smaller than 200MHz and as expected, near the locking region (periodic(P)-mode) the frequency pulling is achieved (quasi-periodic(Q)-mode).[33,110] Indeed, theoretical computations and experiments indicate that the use of a microwave source can give rise to more complex behaviours as listed below. (i) *Hysteretic synchronization*[107,108] where, for a fixed bias current, starting from the Q-mode and increasing/decreasing the frequency of the microwave source towards the phase locking region the P-mode is excited at a frequency $f_P$, after that decreasing/increasing the frequency of the microwave source towards the frequency pulling region the Q mode is excited at a frequency $f_Q$ smaller/larger than $f_P$. It is important to underline that the hysteric synchronization has been achieved only at very low temperature. (ii) *Fractional synchronization*[109] where the locking region is achieved also at integer and fractional ratio between the oscillator and the microwave frequency. Fig. 10 displays the power spectra as function of microwave frequency with the frequency of the excited mode with larger power well identified. As can be observed, the fractional locking is achieved for several non-integer values. From theoretical point of view, it has been demonstrated that to observe fractional synchronization the microwave field should be applied to the free layer in order to break the symmetry of the system. (iii) *Parametric excitation*[118] where the system is biased with a subcritical value of the current, and the persistent magnetization oscillation is excited by a weak microwave field with a frequency two times the one of ferromagnetic resonance frequency. (iv) *Stochastic resonance*[119,120] where in a super-paramagnetic nanomagnet the microwave source can induce synchronized mode hopping among static and oscillatory states. (v) *Phase slip*[121] where the combined action of a microwave current and field (same frequency) for some frequency range in the locking region induce abrupt jumps of $2\pi$ in the oscillator phase giving rise to an additional dissipation mechanism to take into account.

The fundamental understanding of the origin of the non-autonomous behaviour of STNOs will permit to design array of STNOs that can be synchronized by means of a "weak" microwave source.

## Summary and future perspective

The STNOs have now a promising horizon for their potential applications. They are frequency-tuneable by current and magnetic field, small size, low operation voltage and easy to integration on-chip. Although the significant progress in the improvement of the output power and the linewidth, many challenges remain to solve before that STNOs will ever be implement into commercial products. One concern is to remove the need of large external magnetic field. To date the observations of large power microwave emission are only achieved under a large bias field, which is not convenient for on-chip applications. Several ideas have been suggested to solve this issue,[37,40,52,54,62,122] but the "state of the art" solution could be based on the use of pillar-shaped MTJs with a perpendicular free layer and a planar "fixed" layer.[25] In addition to those challenges, engineering issues about device miniaturization, impedance match between STNO and an interconnecting guide, reduction of the drive current density, and increasing of the generation efficiency are also needed to be addressed. It is expected that most of these issues will be solved in the future by developing new magnetic materials and new STNO geometries. For instances, the voltage-driven magnetization dynamics[123,124] may substantially reduce the drive current density, and significantly minimize the device variations due to the avoidance of the very thin barrier layer. Recent works demonstrated the high performance of STNOs by the utilization of the interface perpendicular anisotropy,[22,26,65] and the very promising design of three-terminal STNO exploring the spin transfer torque from the spin-Hall effect.[125] Those works suggest that the voltage-control of the perpendicular anisotropy in combination with spin-Hall effect opens a new way for developing high-power, broad tunability, low-noise devices providing greater and more versatile tuning of the frequency and amplitude of the microwave signal.

## Acknowledgements

We acknowledge financial support from the 100 Talents Programme of The Chinese Academy of Sciences, the National Science Foundation of China (11274343), and the Nanoelectronics Research Initiative (NRI) through the Western Institute of Nanoelectronics (WIN). This work was also supported by Spanish Project under Contract No. MAT2011-28532-C03-01.

†E-mail: zhongming.zeng@gmail.com or gfinocchio@unime.it

11